\documentclass[usenatbib,usedcolum,usegraphicx]{mn2e}

\title[
Extinction Map of the Galactic center: OGLE-II Galactic bulge fields
      ]{
Extinction Map of the Galactic center: OGLE-II Galactic bulge fields
 }

\author[             T. ~Sumi
       ]
       {             Takahiro ~Sumi \\
    Princeton University Observatory, Princeton, NJ 08544-1001, USA,
    e-mail: sumi@astro.princeton.edu 
}
\date{Accepted 
      Received
      in original form }

\pagerange{\pageref{firstpage}--\pageref{lastpage}}

\begin{document}
\maketitle
\label{firstpage}

\begin{abstract}
We present the reddening ($E(V-I)$) and Extinction maps in $V$-band 
($A_V$) and $I$-band ($A_I$) for 48 Optical Gravitational Lensing 
Experiment II (OGLE-II) Galactic bulge (GB) fields, covering a range
of $-11^\circ <l< 11^\circ$, with the total area close to 11 square 
degrees. These measurements are based on two-band photometry of Red 
Clump Giant (RCG) stars in OGLE-II $VI$ maps of GB.  We confirm the
anomalous value of the ratio of total to selective extinction
$R_{VI} \equiv A_V / E(V-I) = 1.9 \sim 2.1$, depending on the line 
of sight, as measured by \cite{uda03}.  By using the average value 
of $R_{VI}=1.964$ with the standard deviation $sdev=0.085$, we 
measured $E(V-I)$, $A_V$ and $A_I$, and we obtained extinction and 
reddening maps with a high spatial resolution of $ 26.7''\sim 106.8''$,
depending on the stellar density of each field.  We assumed that 
average, reddening corrected colours of red clump giants are the same 
in every field. The maps cover the range $0.42<E(V-I)<3.5$, $0.83<A_V<6.9$
and $0.41<A_I<3.4$ mag respectively.  The zero points of these maps are
calibrated by using $V-K$ colours of 20 RR Lyrae ab variables (RRab) 
in Baade's window.  The apparent reddening corrected $I$-band magnitudes
of the RCGs change by $ +0.4 $ mag while the Galactic coordinate $l$ 
varies from $+5^{\circ}$ to $-5^{\circ}$, indicating that these stars 
are in the Galactic Bar. The reddening corrected colour of RRab and 
RCGs in GB are consistent with colours of local stars, while in the 
past these colours were claimed to be different.
\end{abstract}


\section{Introduction}
A study of stellar populations and stellar dynamics in the Galactic Bulge
(GB) is important for understanding how bulges formed, what are their 
populations, gravitational potential and structure.

Several gravitational microlensing survey groups have found hundreds of events 
towards the Galactic center and disc (EROS: \citealt{der99}; OGLE: \citealt{uda00};
\citealt{woz01}; MACHO: \citealt{alc00}; MOA: \citealt{bon01}; \citealt{sum03a}),
and thousands are expected in the upcoming years by MOA
\footnote{{\tt http://www.phys.canterbury.ac.nz/\~{}physib/alert/alert.html}},
OGLE-III
\footnote{ {\tt http://www.astrouw.edu.pl/\~{}ogle/ogle3/ews/ews.html}}
and other collaborations.
The data from such microlensing surveys is useful to 
study the Galactic structure by measuring the microlensing
optical depth (\citealt{uda94}; \citealt{alc97,alc00}; \citealt{sum03a};
\citealt{afo03}; \citealt{pop03};) and the proper motions of stars
(\citealt{sum03,sum03b}), and well suited for numerous other scientific projects
(see \citealt{pac96}; \citealt{gou96}).

However, as is well known, the extinction due to the dust is very significant
towards the GB. This affects the Color
Magnitude Diagram (CMD) of the field and makes a separation of stellar 
populations difficult.  To correct for these effects the measurements of
the extinction in these fields are crucial.

\cite{schl98} made all sky extinction map using COBE/DIRBE data, 
which overestimate the extinction towards the GB because of background dust 
(\citealt{dut03}).  \cite{schu99} and \cite{dut03} constructed
$K$-band extinction maps of the Galactic
central region with a resolution of $4'$ by using $J$ and $K$ photometry of 
the upper giant branch stars in DENIS and 2MASS data respectively. 
Some determinations of the extinction towards Baade's Window have
been performed with a number of different techniques,
including those of stellar simulation (\citealt{ng96}), mean magnitudes of 
red-clump stars (\citealt{kir97}), the absolute magnitude of RR Lyrae stars
(\citealt{alc98b}) and magnitude of the K-giants (\citealt{goupop98}).
The Large-Scale Extinction Map of the Galactic Bulge was made by using
the mean colour of all stars in the MACHO Project Photometry (\citealt{popcb03})

\cite{woz96} proposed a method to investigate the ratio of total to selective extinction
based on two-band photometry of Red Clump Giants (RCGs).
The RCGs are the equivalent of the horizontal branch stars for a metal-rich 
population, i.e., relatively low-mass core helium burning stars.  RCGs in the 
Galactic bulge occupy a distinct region in the colour magnitude diagram 
(\citealt{sta00} and references therein). The intrinsic width of the luminosity and 
colour distribution of RCGs in the Galactic bulge is small, about 0.2 mag
(\citealt{sta97}; \citealt{pac98}).

The CMD is used to obtain the quantitative
values of the offset on the CMD between the different subfields, caused
by differential extinction. They used RCG-dominated parts of the CMDs for
determining the offsets, the clump being seen at fainter magnitudes and
redder colours in subfields with higher extinction.  They applied this method
to the OGLE-I data and then found the  ratio of total to selective extinction 
$R_{VI}\equiv A_V/E(V-I) = 2.44$.  This is consistent with \cite{ng96}.
\cite{sta96} applied this method to the OGLE-I data to obtain differential
extinction $A_V$ and reddening $E(V-I)$ in a $40' \times 40'$ region of 
Baade's window, with resolution of $\sim 30''$. They estimated
$R_{VI} = 2.49\pm 0.02$.
\cite{pac99} and  \citealt{sum03a} applied this method to OGLE-II 
($14'.2\times 14'.2$ with resolution of $20''\times 20''$) and MOA 
($16$ deg$^2$ with resolution of $3.45'\times 3.45'$) data respectively. 
They first made a reddening map for their fields because determining the
reddening $E(V-I)$ (horizontal shift in the CMD) is easier than $A_V$ and $A_I$
(vertical shift in the CMD).  Then the extinction map was calculated
according to the following formulae:

\begin{eqnarray}
  \label{eq:ai} 
  A_I &=&  R_I \times E(V-I), \\
  \label{eq:av} 
  A_V &=&  R_{VI} \times E(V-I).
\end{eqnarray}
with "standard" values of $R_{VI}=2.5$ and  $R_{I}\equiv A_I/E(V-I) = 1.5$.

\cite{pac98} and \cite{stu99} found that the mean 
$V-I$ colours of GB RCGs and RR Lyrae, dereddened with Stanek's map,
are redder than colours of their nearby counterparts. \cite{pop00} summarized
possible explanations of these discrepancies, and noted that the simplest 
and the most plausible 
explanation is a non-standard interstellar extinction.
The discrepancy would vanish if $R_{VI}=2.1$  rather than the standard value of
$2.5$.  \cite{uda03} showed that there is indeed an anomaly in the extinction
law, with $R_{VI}=1.9 \sim 2.3$, depending on the direction of the line of 
sight.

In this paper we  confirm the anomalous value of $R_{VI}$, and  by using new
value, we construct extinction maps for 48 Galactic Bulge
fields observed by the Optical Gravitational Experiment
\footnote{see {\tt http://www.astrouw.edu.pl/\~{}ogle} or 
{\tt  http://bulge.princeton.edu/\~{}ogle}}
 II (OGLE-II; \citealt{uda00}).

In \S\,\ref{sec:data} we describe the data.  
We measure the reddening line in \S\,\ref{sec:AIEVI}.
We construct the reddening and extinction maps in \S\,\ref{sec:extinctionmap}.
Discussion and conclusion are given in \S\,\ref{sec:disc}.

\section{DATA}

\label{sec:data}

We use the $VI$ photometric maps of standard OGLE template (\citealt{uda02}),
which contain $VI$ photometry and astrometry of $\sim 30$ million stars in the 
49 GB fields. Positions of these fields (BUL\_SC1 $\sim$ 49) can be seen in 
\cite{uda02}. We do not use BUL\_SC44 in this work because most of RCGs in
this field are close to, or even below the $V$-band detection limit of OGLE
due to high extinction. 
The photometry is the mean photometry from a few hundred measurements in the $I$-band
and several measurements in $V$-band collected during the second phase of the
OGLE experiment between 1997 and 2000. Accuracy of the zero points of photometry
is about 0.04 mag. A single 2048 $\times$ 8192 pixel frame covers an area of 
0.24 $\times$ 0.95 deg$^2$ with pixel size of 0.417 arcsec/pixel.  Details of the
instrumentation setup can be found in \cite{uda97}.

\section{The ratio of total to selective extinction}
\label{sec:AIEVI}

To measure the ratio of total to selective extinction, i.e. $R_{I} \equiv 
A_I/E(V-I)$,
we make use of the position of RCGs in the  ($I$,$V-I$) CMD, as it was done
by \cite{uda03}, but
contrary to \cite{woz96} and \cite{sta96}, who used the  ($V$,$V-I$) CMD.
The reddening corrected $I$-band magnitude of the RCGs does not vary with
colour, while the $V$-band magnitude is a function of $V-I$ (\citealt{pac98}),
and using it can lead to systematic errors (\citealt{pop00}).
We make a preliminary assumption that the average, reddening corrected RCG
colour is constant, $\langle V-I \rangle _{0,RC} = 1.0$, which is 
approximately the colour of nearby RCGs as measured by Hipparcos
(\citealt{pac98}).
The mean colour $\langle V-I \rangle_{\rm RC}$ and $I$-band magnitude
$\langle I \rangle_{\rm RC}$ of RCGs follow reddening line with a slope 
$R_{I}$ and a constant $I_{0, \rm RC}$, to be determined for every field

\begin{equation}
\langle I \rangle_{\rm RC}=
I_{0,\rm RC}+R_I
( \langle V-I \rangle_{\rm RC} - \langle V-I \rangle _{0,RC} ).
\label{eq:reddningcurve}
\end{equation}

In this Section, we measure the slope $R_I$ and brightness of 
RCGs $I_{0,RC}$ in Eq.\ref{eq:reddningcurve}, with a low spacial
resolution, but with a significant number of RCGs. In the next 
Section, we measure the colour of RCGs
with high spacial resolution by using these $R_I$ and $I_{0,RC}$.
This is because an identification of RCG centroids in $I$
is more difficult than in $V-I$ because of the vertical structure
of Red Giants overlapping RCGs in the CMD.

In this Section, there are 2 steps to find the RCG centroids in
the CMD.  At a first step, we divide the field into "bins" to take
spacial differences of the extinctions into account, then we measure 
the rough mean colour of RCGs in each bin. At a second step, we 
combine bins with similar RCG colour into group to enlarge the 
significance of RCGs, then we estimate the RCG centroids by the 
Gaussian fit.

As a first step,
we divide each field into $16 \times 64 = 512$ "bins" which have 
$128 \times128$ pixels each.  As these "bins" are relatively large
there may be a considerable differential reddening within them,
and the RCG may be elongated along the reddening line.  

We select an
elongated window within which the RCGs are located, following the
reddening line as given by the Eq. (\ref{eq:reddningcurve}) with a 
width of $\pm 0.5$ mag in $I$.  As we do
not know the correct reddening slope and the correct magnitude of RCGs,
we adopt a broad range of trial values: $ 0.6 < R_I < 2.0 $, and
$ 14 < I_{0,RC} < 15 $.  We also adopt a broad range of colours for
the search of RCGs, selected to be within $ 1.4 < (V-I) < 5 $; 
the boundaries are adjusted for each field to minimize the contamination
by blue disk main sequence stars and faint red bulge main sequence stars.
We measure the average $\langle V-I \rangle_{\rm RC}$ in each bin
using 2-$\sigma$ clipping. These values are used as the initial values of
the RCGs colour in the next paragraph.

In a given bin we measure the average colour and magnitude of RCGs
within the circle with a radius of 0.4 mag in the CMD centered at  
the colour $\langle V-I \rangle_{\rm RC}$ on the reddening line. 
These new average values:
$\langle V-I \rangle_{\rm RC}$ and $\langle I \rangle_{\rm RC}$,
calculated for all $ 128 \times 128 $ pixel bins, are used to obtain an 
improved value of $ R_I $ and $ I_{0,RC}$. This process is iterated 
until the values of $ R_I $ and $ I_{0,RC}$ do not change any more.  
We found that this final value for a given OGLE field was independent of 
the first guess in the ranges $0.6< R_I <2.0$ and $14< I_{0,RC}<15$,
and are roughly the same as the more precise values measured in the second step.

Even if we locate the window slightly higher or lower, i.e. $I_{0,\rm RC}$
in Eq.(\ref{eq:reddningcurve}), than the true RCG centroid in $I$, 
we would get roughly same colour as the true RCGs colour 
because the RCGs colour are very similar to the colours of red giants
which are somewhat brighter or fainter than RCGs.
We use only resultant average colour $\langle V-I \rangle_{\rm RC}$ of 
each bin in the following analysis, to arrange bins in order of their 
extinction in a given field. So, as long as we fix the value of $ R_I $
and $ I_{0,RC}$ in a given field these colours can be used to arrange 
the bins.

As a second step we arrange the bins in a given field in order of 
extinction by using $\langle V-I \rangle_{\rm RC}$. Then we 
combined these bins into groups from low extinction to high extinction until 
each group is filled by $\sim 1000$ RCGs. Given a large number of RCGs 
in each group of bins we could find the RCGs positions in the CMD for
each group independently. We measured RCGs centroid of each group in the
CMD by following three methods:

[1] In a given  group of bins we measure the average colour and magnitude of 
RCGs within the circle with a radius of 0.3 mag in the CMD, centered at the
initial values of colour
and magnitude of RCGs.  These new average values are used for new RCG selection. 
This process is iterated until the value  $\langle V-I \rangle_{\rm RC}$ and
$\langle I \rangle_{\rm RC}$  do not change any more.
We found these final values do not depend on their initial values which can 
be given by the rough positions of RCGs in previous section, as long as their
real positions are within this initial circle.

[2] Similar with [1] but  with a larger radius of 0.45 mag and weighting
$V-I$ and $I$ with two dimensional Gaussian with $\sigma= 0.15$ mag.

[3] We fit the distribution of stars in each group
with the power law plus Gaussian luminosity function:
\begin{equation}
  \phi_I(I) = p_0 10^{p_1 I} +  p_2 \exp\left( -\frac{(I-\langle I\rangle_{\rm RC})^2}{2 \sigma_{I,\rm RC}^2} \right),
\label{eq:powgauss}
\end{equation}
where $p_0$, $p_1$, $p_2$ and $\sigma_{I,\rm RC}$ are free parameters,
calculated for each group of bins separately.
RCGs are selected within the circle with radius of $0.4$ mag centered at 
this best fit $\langle I \rangle_{\rm RC}$ and $\langle V-I\rangle_{\rm RC}$ 
measured by method [1]. We fitted the colour distribution of selected RCGs 
with another Gaussian: 
\begin{equation}
  \phi_{VI}(V-I) = p_3 \exp\left( -\frac{\left((V-I)-\langle V-I\rangle_{\rm RC}\right)^2}{2 \sigma_{VI,\rm RC}^2} \right),
\label{eq:gauss}
\end{equation}
where $p_3$ and $\sigma_{VI,\rm RC}$  are free parameters.

\begin{figure}
\begin{center}
\includegraphics[angle=-90,scale=0.35,keepaspectratio]{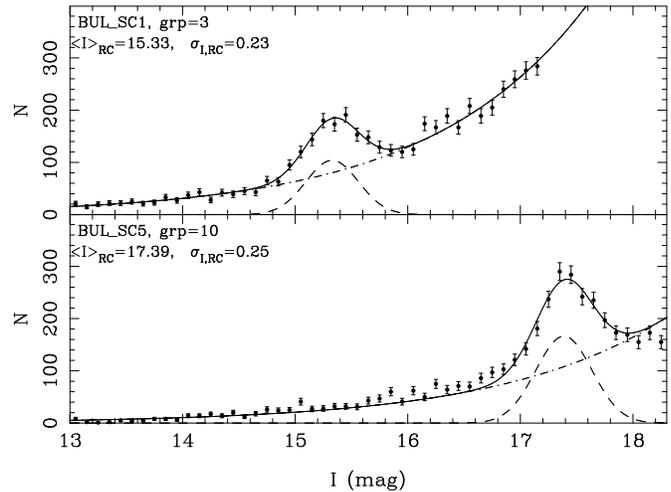}
\caption{
Sample $I$-band Luminosity Functions of stars in relatively low 
(upper panel: BUL\_SC1, grp=3) and high (lower panel: BUL\_SC5, grp=10) 
extinction fields.  Solid lines are the best fits by equation 
(\ref{eq:powgauss}).
Dashed and dot-dashed lines represent the Gaussian and power law components 
in the best fits.
  \label{fig:LF}
  }
\end{center}
\end{figure}

In Fig. \ref{fig:LF} we show the sample $I$-band Luminosity Functions 
of stars in a relatively low (upper panel: BUL\_SC1, grp=3)  
and high (lower panel: BUL\_SC5, grp=10) extinction fields with the best fit
by equation (\ref{eq:powgauss}). Here groups (grp) which have roughly 1,000 
RCGs are numbered from 1 in order of increasing extinction. One can see 
how significantly we can measure the RCGs centroids in these large groups.

We arrange 44 OGLE-II fields into 11 regions of fields close to each other,
as presented in table \ref{tbl:RI}.
Regions A, B, C and D are identical with those in \cite{uda03}.
Fields BUL\_SC6, 7, 47, 48, 49 are not analyzed because there are 
too few RCGs in these fields and very little differential extinction.
We assume that the slope of the reddening line, $R_I$, is the same 
within each region, but it may be different in different regions.
For every region we had a collection of groups of bins, and the values
of centroids of the RCGs:
($ \langle I \rangle _{RC} , \langle V-I \rangle _{RC}) $ for each
group.  We fitted these data with Eq.(\ref{eq:reddningcurve}).
All $R_I$ measured by different methods are consistent with
each other within their errors. In the following analysis we use 
the results by method [3] because they have the smallest scatter
in the fitting.

\begin{figure}
\begin{center}
\includegraphics[angle=0,scale=0.47,keepaspectratio]{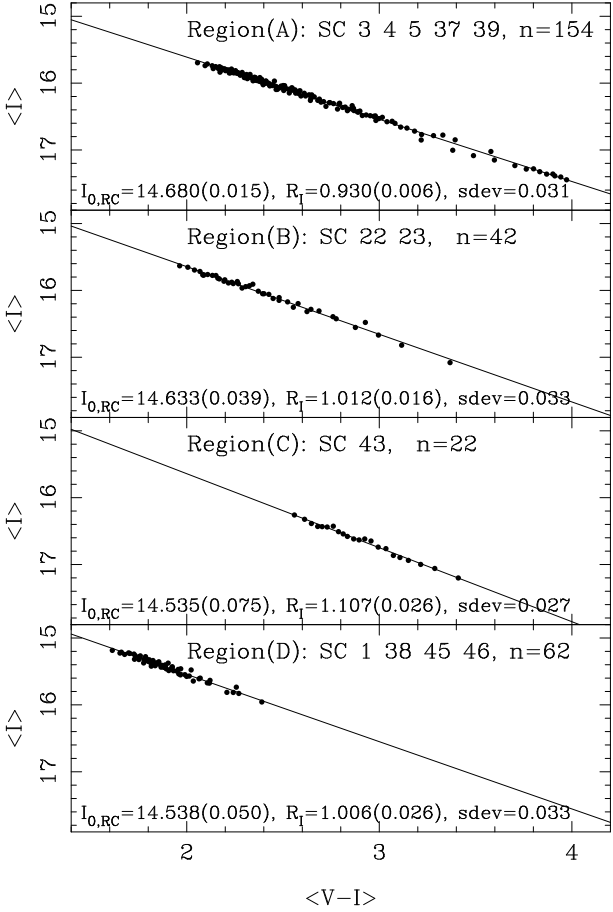}
\caption{
Centroids (filled circle) of RCGs in the ($\langle I \rangle$, $\langle V-I \rangle$) 
CMD in the region (A), (B), (C) and (D) (from top to bottom) measured with
method [3], as described in the text.
The best fit line, parameters and the standard deviation ($sdev$)
are also given in the figure.
  \label{fig:IVI}
  }
\end{center}
\end{figure}

In Fig. \ref{fig:IVI} we show the distribution of centroids for regions
(A), (B), (C) and (D) 
with the best fit lines.  The best fit value $R_I$, the error 
$\sigma$, and the standard deviation $sdev$ for all regions, and their average 
values are given in table \ref{tbl:RI}.  We also give the the mean $R_I$ for
all regions except (A), because region (A) has a significantly smaller $R_I$ 
than other fields, and it has a
very small error.  In region (A) (Top panel of Fig. \ref{fig:IVI}), 
we did not use groups with $\langle V-I\rangle_{\rm RC}>4$ because they are 
close to detection limit in $V$-band, and 
their center in the CMD is shifted systematically to 
brighter $I$. We believe that groups with $\langle V-I\rangle_{\rm RC}<4$
are not affected by this effect because $R_I$ does not change when
this limit is reduced to smaller values. 

\begin{figure}
\begin{center}
\includegraphics[angle=0,scale=0.47,keepaspectratio]{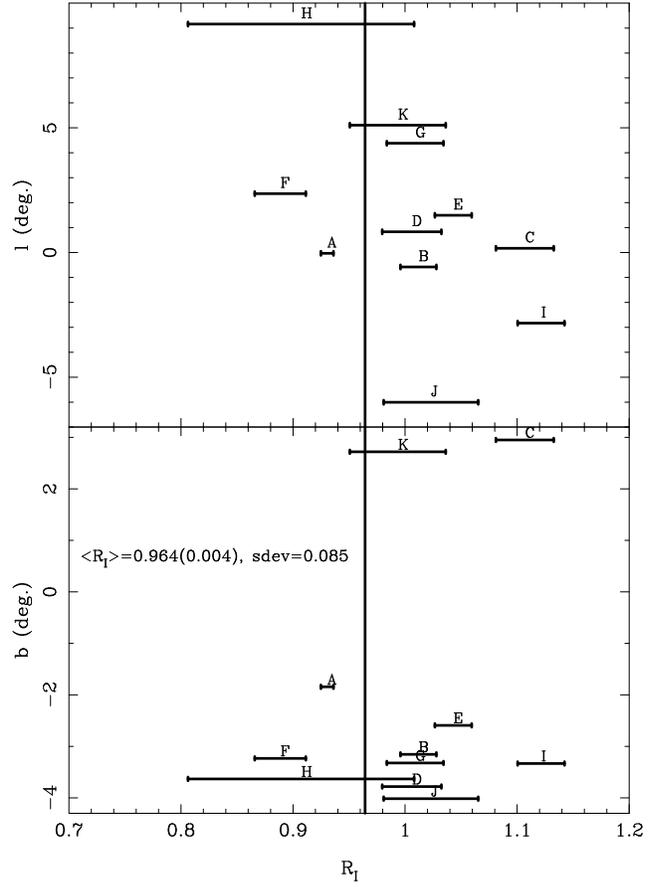}
\caption{
The ratio of total to selective extinction $R_I\equiv A_I/E(V-I)$ 
for each region as a function of Galactic coordinates.  The values of $R_I$
correspond to the middle of the error bars, and the average
$\langle R_I \rangle$ is shown with a vertical line.
  \label{fig:RIlb}
  }
\end{center}
\end{figure}

Our values of $R_I$ are consistent with \cite{uda03} and are significantly
smaller than the standard reddening law ($\sim 1.5$). The measured 
$R_I$ are shown as a function of the Galactic longitude ($l$) and 
latitude ($b$) in Fig. \ref{fig:RIlb}. We can see that $R_I$ is slightly
different from region to region, but there is no strong systematic 
dependence on the Galactic coordinates $l$ or $b$.  There might be a 
weak dependence in $l$, $dR_I/dl=-0.0102\pm0.0049$.  This trend gives the
 largest difference in $R_I$ of 0.08 in regions (H) and (J).
We think that this trend can be neglected in the following analysis 
because we take the scatter of 0.085 in $R_I$ into account 
during the error estimation in reddening and extinctions.

\begin{figure}
\begin{center}
\includegraphics[angle=-90,scale=0.35,keepaspectratio]{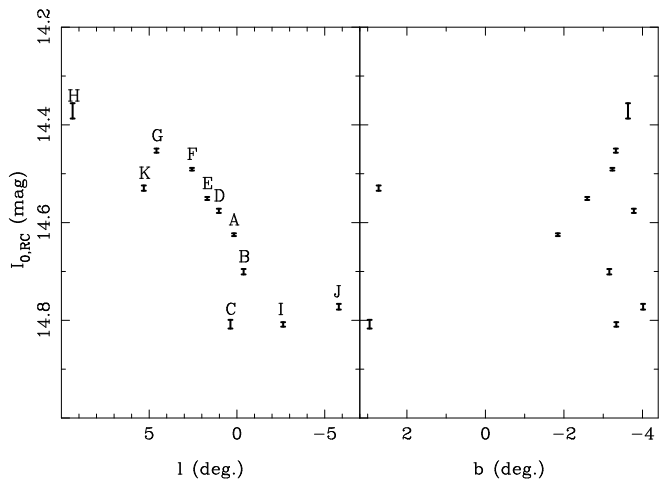}
\caption{
The constant $I_{0,\rm RC}$ (cf. Eq. (\ref{eq:reddningcurve})) is shown for each 
region as a function of Galactic coordinates adopting a fixed value $R_I=0.964$.
The values correspond to the middle of error bars.  Note region (C) and (K) 
are fields with $b >0$. 
  \label{fig:I0lb}
  }
\end{center}
\end{figure}

In the following analysis we use the mean value of the reddening slope
for all regions i.e. $R_{I}=0.964$ with $sdev = 0.085$.  The values of
$I_{0,\rm RC}$, obtained by fitting Eq. (\ref{eq:reddningcurve}) 
with the fixed slope $R_{I}=0.964$ for each region, are given in 
table \ref{tbl:I0} and are plotted in Fig. \ref{fig:I0lb} as a function
of Galactic coordinates $l$ and $b$.  A clear evidence of the bar is 
apparent between $-4<l<4$ in the upper panel of this figure, which is 
consistent with \cite{sta97}. Note that regions (C) and (K) are shifted with
respect to the pattern indicated by regions (G), (F), (E), (D), (A), (B);
they are on the other side of the Galactic plane, i.e. in $b>0$.

\begin{table*}
 \centering
 \caption{The ratio of total to selective extinction $R_I$ for each region. 
    \label{tbl:RI}}
    \begin{tabular}{clccccc}\\
    Region &fields (BUL\_SC)& $l (deg.)$ & $b (deg.)$ & 
    $R_{I}$ & $\sigma_{R_{I}}$ & $sdev$

\\
  \hline
A & 3 4 5 37 39 & 0.168 & -1.844 &  0.930 & 0.006 & 0.031  \\
B & 22 23 & -0.380 & -3.155 &  1.012 & 0.016 & 0.033  \\
C & 43 & 0.370 & 2.950 &  1.107 & 0.026 & 0.027  \\
D & 1 38 45 46 & 1.030 & -3.780 &  1.006 & 0.026 & 0.033  \\
E & 20 21 30 34 & 1.692 & -2.593 &  1.043 & 0.016 & 0.028  \\
F & 2 31 32 33 35 36 & 2.560 & -3.233 &  0.889 & 0.023 & 0.027  \\
G & 16 17 18 19 42 & 4.582 & -3.322 &  1.009 & 0.025 & 0.032  \\
H & 8 9 10 11 12 13 & 9.360 & -3.632 &  0.907 & 0.101 & 0.078  \\
I & 24 25 40 41 & -2.632 & -3.333 &  1.121 & 0.021 & 0.031  \\
J & 26 27 28 29 & -5.805 & -4.015 &  1.023 & 0.042 & 0.036  \\
K & 14 15 & 5.305 & 2.720 &  0.994 & 0.043 & 0.026  \\
All & - &  -  & - & 0.964 & 0.004 & 0.085 \\
without A & - &  -  & - & 1.026 & 0.008 & 0.075 \\

\end{tabular}
\end{table*}

\begin{table}
 \centering
 \caption{Extinction corrected $I$ magnitude of RCGs $I_{0,RC}$ measured with fixing $R_I = 0.964$ for each region.
    \label{tbl:I0}}
    \begin{tabular}{clcccccc}\\
    Region & $I_{0,RC}$ & $\sigma_{I_{0,RC}}$ & $sdev$ &
             $I_{0,RC}^{*}$ & $\sigma_{I_{0,RC}^{*}}$ &
       \\
  \hline
A & 14.625  & 0.003 & 0.034 &     14.652 & 0.024 \\
B & 14.700  & 0.005 & 0.036 &     14.728 & 0.024 \\
C & 14.808  & 0.009 & 0.042 &     14.835 & 0.025 \\
D & 14.576  & 0.004 & 0.034 &     14.603 & 0.024 \\
E & 14.551  & 0.003 & 0.031 &     14.578 & 0.024 \\
F & 14.491  & 0.003 & 0.028 &     14.518 & 0.024 \\
G & 14.453  & 0.004 & 0.033 &     14.480 & 0.024 \\
H & 14.371  & 0.016 & 0.077 &     14.398 & 0.028 \\
I & 14.808  & 0.005 & 0.041 &     14.836 & 0.024 \\
J & 14.772  & 0.006 & 0.036 &     14.799 & 0.024 \\
K & 14.529  & 0.005 & 0.026 &     14.557 & 0.024 \\
\end{tabular}
Note: $I_{0,RC}^{*}$ represent the value after the zero-point correction.
\end{table}

\section{Extinction Map}
\label{sec:extinctionmap}
\subsection{Relative Reddening}
\label{sec:relativeEVI}

In this section we estimate the mean RCGs colours for each bin 
$\langle V-I \rangle_{\rm RC}$, and we transform them to the relative reddening,
$E(V-I)_{\rm RC}=\langle V-I \rangle_{\rm RC}-\langle V-I \rangle_{0,\rm RC}$,
assuming the intrinsic colour of RCGs $\langle V-I \rangle_{0,\rm RC}=1.0$.  
Then we estimate the relative extinctions in $V$-band ($A_{V,\rm RC}$) and 
$I$-band ($A_{I,\rm RC}$) by Eq. (\ref{eq:av}) and (\ref{eq:ai}).

In the following analysis we adopt the reddening line with the
slope $ R_I = 0.964 $ and a constant $I_{0,RC}$ as given in table \ref{tbl:I0}.
Thanks to fixing $ R_I$ and $I_{0,RC}$, we can get accurate
$\langle V-I \rangle_{\rm RC}$ with a smaller number of RCGs, i.e.
with high resolution in space (small bin) and in reddening (small group).
Furthermore we introduce a new indicator $\langle V-I \rangle_{\rm all}$ 
which is the average colour of all stars in each bin, 
to represents the level of extinction in each bin to arrange bins into group. 
Because the number of all stars are much larger than RCGs, we 
to get higher spacial resolution.

\begin{figure}
\includegraphics[angle=0,scale=0.45,keepaspectratio]{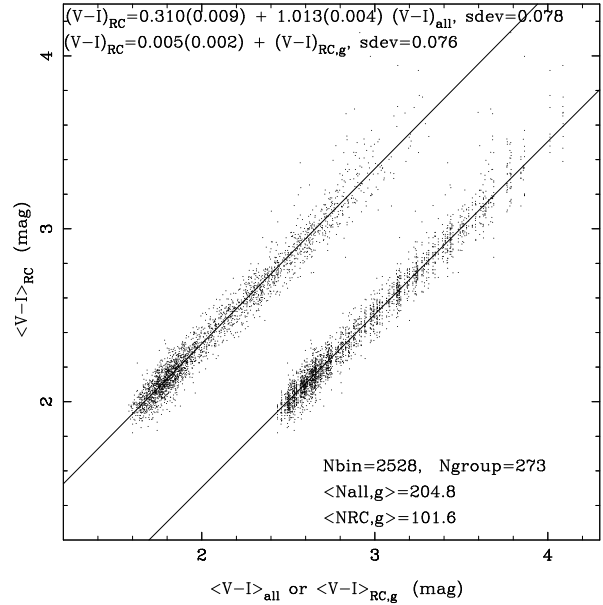}
\begin{center}
\caption{
The mean colour of RCGs  $(V-I)_{\rm RC}$ in each bin as a function of the mean 
colour of all stars in the bin ($(V-I)_{\rm all}$, upper sequence) and the
mean colour of RCGs in each group ($(V-I)_{\rm RC,g}$, lower sequence) in the field BUL\_SC22.  
$(V-I)_{\rm RC,g}$ is shifted by +0.5 mag for clarity.
\label{fig:VIrc-VIg}
}
\end{center}
\end{figure}

We divide each field into a new set of small "bins" with the size 
in the range $64\times64$ to $157\times157$ pixels, chosen so that each bin 
has $\sim 200$ stars. Next, we measure the average colour of all stars 
in each bin, $\langle V-I \rangle_{\rm all}$.  We also measure the average
colour of RCGs in each "bin", $\langle V-I \rangle_{\rm RC}$,
following method [1] described in the previous section, with a radius of 
0.5 mag, allowing the center of RCGs to lie only on the reddening line.
The initial values of $\langle V-I \rangle_{\rm RC}$ for this method
are estimated by using the parallelogram along this reddening line.
This may be fairly uncertain statistically as there are only several RCGs per
bin, but may suffer from less systematics than $\langle V-I \rangle_{\rm all}$.
We compare the two sets of colours in Fig. \ref{fig:VIrc-VIg}
(upper sequence) for the field BUL\_SC22.
There is good correlation between $\langle V-I \rangle_{\rm RC}$ and 
$\langle V-I \rangle_{\rm all}$.
Other fields have similar trends, but the slops are not always $1$.
This good correlation imply that $\langle V-I \rangle_{\rm all}$
can be a good extinction indicator, as suggested by \cite{popcb03}.

We arrange bins in a given field in order of extinction by using 
$\langle V-I \rangle_{\rm all}$. Then we combine these bins into 
groups from low extinction to high extinction until
each group is filled by $\sim 100$ RCGs. Note that groups in 
\S\,\ref{sec:AIEVI} have $\sim 1,000$ RCGs.

\begin{figure}
\includegraphics[angle=0,scale=0.5,keepaspectratio]{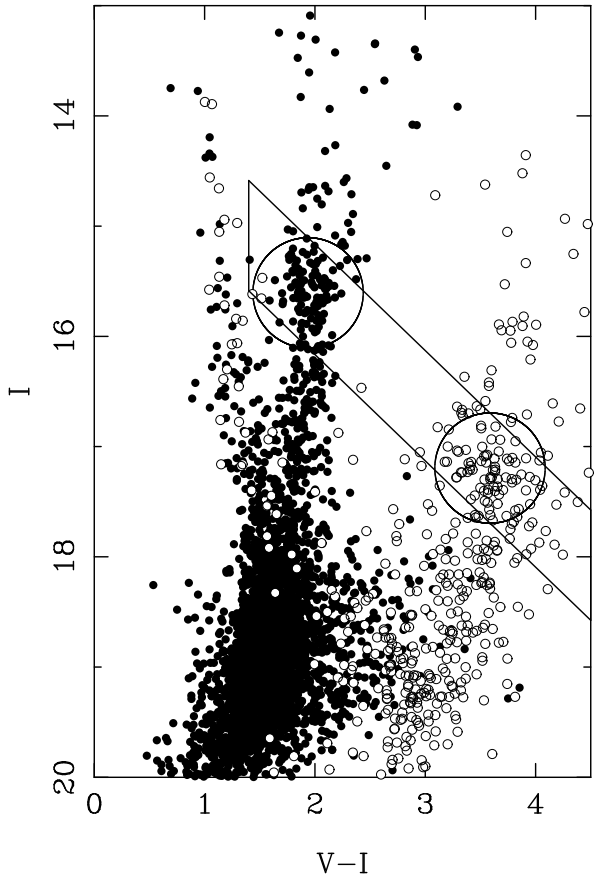}
\begin{center}
\caption{
The CMDs of two groups in BUL\_SC22. Filled and open circles represent stars 
in a group with low and high extinction, respectively. The elongated 
parallelogram represents the window used for the initial selection 
of RCGs.  Two large circles indicate the RCG selection for the two groups.
\label{fig:CMD}
}
\end{center}
\end{figure}

In Fig. \ref{fig:CMD} we show an example of a CMD for two groups of bins in
BUL\_SC22, one with low extinction (filled circles) and one with high 
extinction (open circles).  We calculate the average colour of RCGs
$\langle V-I \rangle_{\rm RC,g}$ for each group of bins using the
method described in previous paragraph. As described above, thanks to fixing
$ R_I$ and $I_{0,RC}$, we can get accurate
$\langle V-I \rangle_{\rm RC, g}$ because the RCGs colour are very similar
to the colours of red giants which are somewhat brighter or fainter than RCGs.

There is a very good correlation
between these colours and $\langle V-I \rangle_{\rm RC}$ obtained for
single bins, as shown in Fig. \ref{fig:VIrc-VIg} (lower sequence).
In this figure the typical width of $\langle V-I \rangle_{\rm RC}$ of bins
in each group can be seen by the gaps of $\langle V-I \rangle_{\rm RC,g}$,
about $<0.1$ mag depending on the number density of stars.
We adopt $\langle V-I \rangle_{\rm RC,g}$ as the mean colour of 
RCGs for bins in this group, because these are based on the large 
statistics of RCGs.

Some bins were not included in any group for variety of reasons:
(1) a bin has fewer than 20 stars, 
(2) a bin has a very bright blue star (mimics low extinction),
(3) a bin has a very bright red star (mimics high extinction).
(4) because of a very high extinction
a large fraction of Bulge red giants is below the detection limit,
so these bins can not be grouped properly 
with $\langle V-I \rangle_{\rm all}$ (in fields BUL\_SC5 and BUL\_SC37).
If there were more than 5 RCGs in a given bin, then we adopted the average
colour of these RCGs $\langle V-I \rangle_{\rm RC}$, which were estimated
in the third paragraph of this section, as the mean colour of 
RCGs for this bin. 
Otherwise these bins are filled by the average value of the neighbours.
$\langle V-I \rangle_{\rm RC}$ in each bin can not be used for the case (3), 
because stars reddened by bright red stars contaminate the RCG region 
in the CMD.

Problematic bins are flagged with "-1" (1), "-2" (2), "-3" (3) and "-4" (4).
If $\langle V-I \rangle_{\rm RC}$ is adopted in a bin then this bin is flagged
with the above value plus "-10".

We do not have confidence in measurements of  
$\langle V-I \rangle_{\rm RC}$ at  $\langle V-I \rangle _{\rm RC} > 4$,
where the detection limit makes RCG centroid in the CMD
systematically brighter and bluer.
This implies that for $\langle V-I \rangle _{\rm RC} > 4$ mag
our maps give a lower limit to the extinction.
We added "-20" to the value of the  flag of such bins.
Only BUL\_SC5 and BUL\_SC37 suffer from this effect.

\subsection{Zero point}

\begin{figure}
\includegraphics[angle=-90,scale=0.37,keepaspectratio]{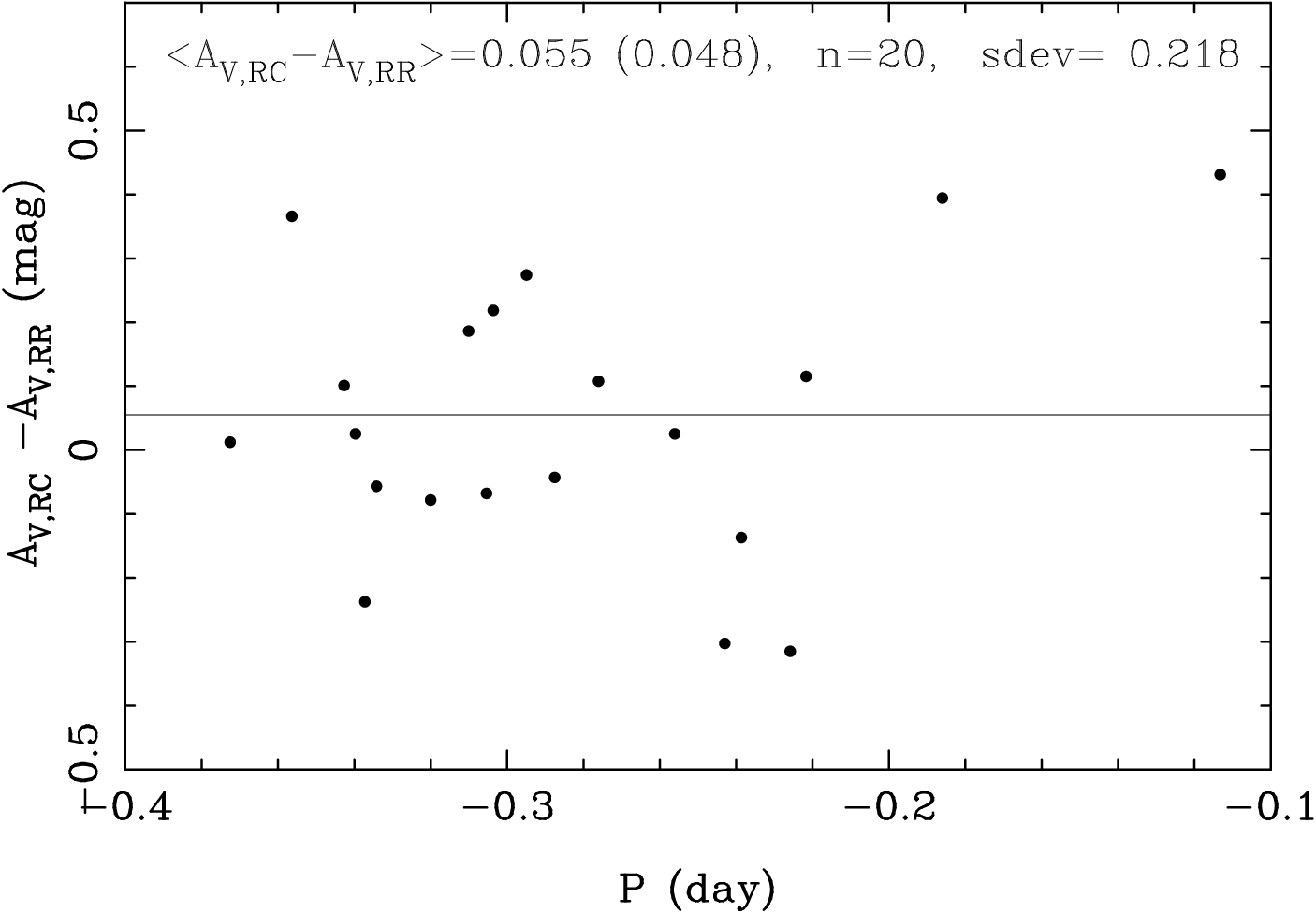}
\begin{center}
\caption{
Difference of relative extinction $A_{V,\rm RC}$ from RCGs (see \S \ref{sec:relativeEVI})
 and $V$-band extinction from 20 RRab in Alcock et al. (1998b),
$A_{V, \rm RR}$, i.e., the same figure as Fig. 3 of  Alcock et al. (1998b).
\label{fig:AV0}
}
\end{center}
\end{figure}

We assume that the average colour of RCG stars, corrected for interstellar
reddening, is the same in every field, and we ignore a possible weak 
dependence on metalicity (\citealt{pac98}).
The zero-point of $\langle V-I \rangle _{\rm 0,RC}$ is calibrated
following \cite{alc98b}.
In Fig. \ref{fig:AV0}, we show the offset of our relative extinction
$A_{V,\rm RC}$ and $V$-band absolute
extinction for 20 RRab in \cite{alc98b}, $A_{V, \rm RR}$.  The average offset 
is $ \langle A_{V,\rm RC} - A_{V, \rm RR} \rangle  = 0.055 \pm 0.048$ 
with the standard deviation of $sdev=0.22$.  The errors and $sdev$ are the 
same as in \cite{alc98b} with \cite{sta96}'s map.
This means that reddening corrected $\langle V-I \rangle$ colour and  
$\langle I \rangle$ magnitude of RCGs
are given as $1+0.055/R_{VI}=1.028$ and the value of $I_{0,\rm RC}$ in table 
\ref{tbl:I0} plus $0.055R_I/R_{VI}=0.027$, respectively.
The extinction corrected $I$-band magnitudes $I_{0,\rm RC}^{*}$ after
zero-point correction are shown in table \ref{tbl:I0}.

\begin{figure*}
\includegraphics[angle=0,scale=1.03,keepaspectratio]{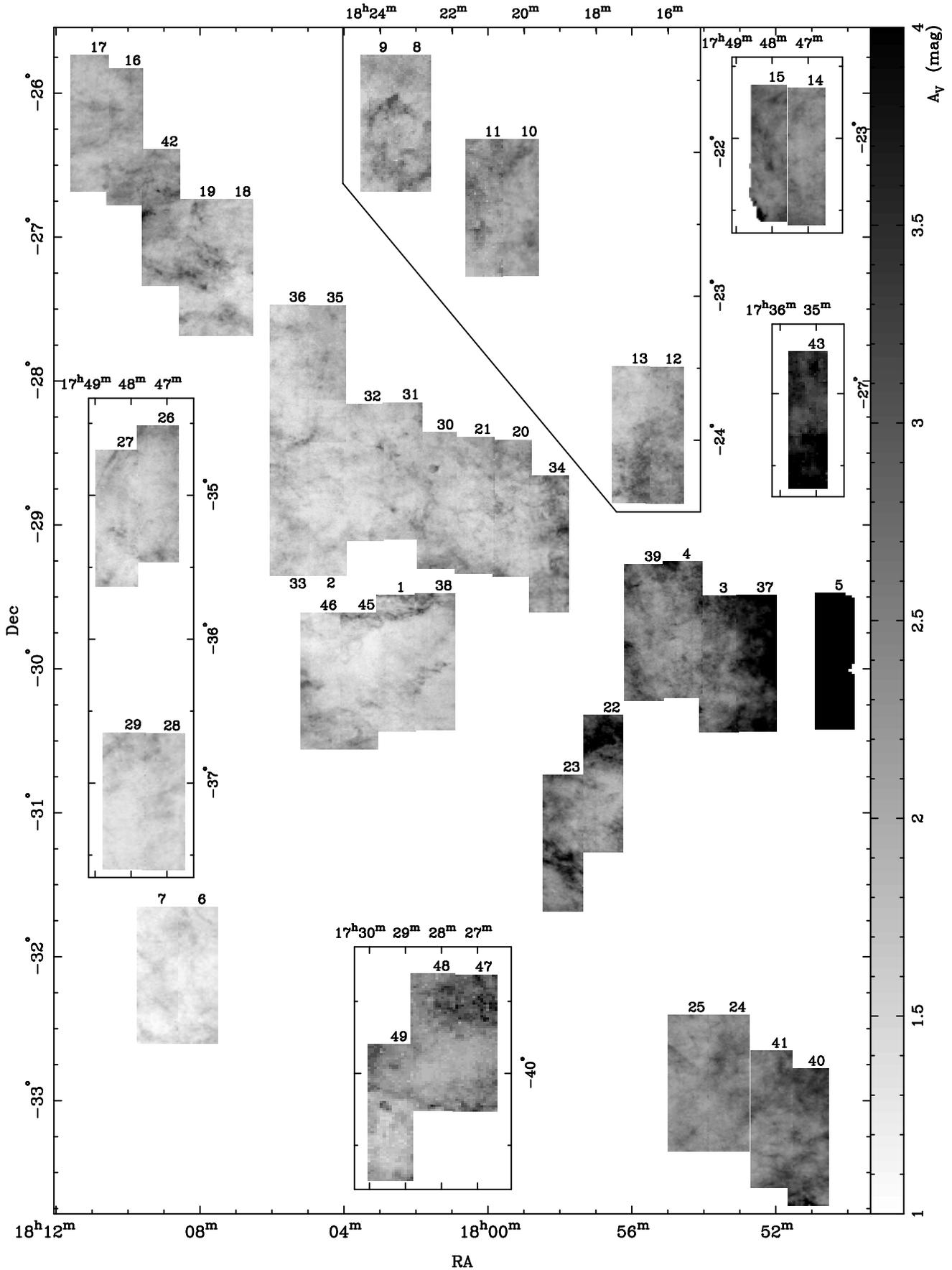}
\begin{center}
\caption{
$A_V$ Extinction map. Distant fields are in small boxes. Gray scale is shown at the right corner.
\label{fig:AVmap}
}
\end{center}
\end{figure*}

\begin{figure}
\includegraphics[angle=0,scale=0.45,keepaspectratio]{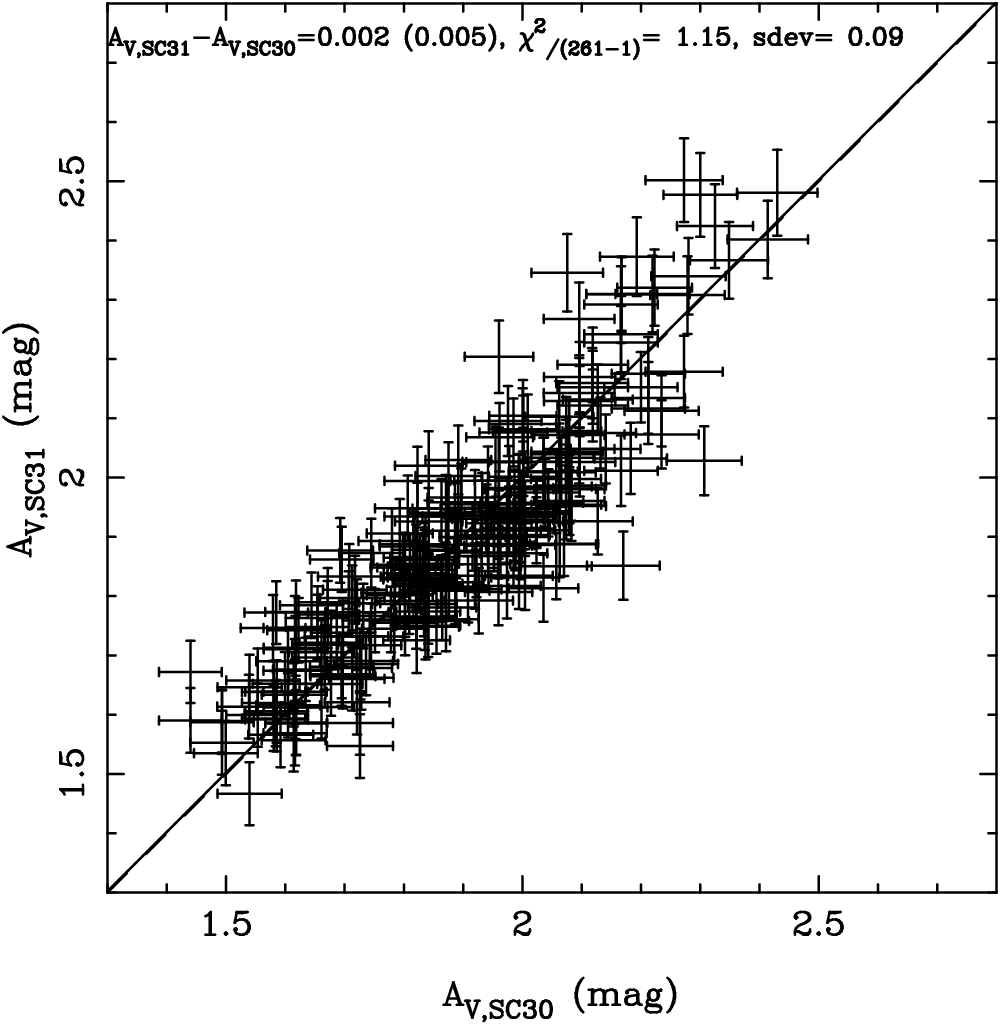}
\begin{center}
\caption{
Comparison of $A_V$ of overlap region in BUL\_SC30 and in BUL\_SC31.
\label{fig:AVSC30-31}
}
\end{center}
\end{figure}

Our extinction and reddening maps are calibrated by this offset.
We show the $A_V$ extinction maps in Fig. \ref{fig:AVmap}.  The
parameters: the bin-size, the number of groups, the total number of all stars,
the average number of all stars in each bin, the total number of RCGs, the
average number of RCGs in each group,
the average values and errors in $E(V-I)$, $A_V$ and $A_I$ are given
in Table \ref{tbl:AVmaps}.  The values
of $\langle \sigma \rangle$ are the statistical errors in relative maps, and
they do not
include zero-point errors $\sigma_{E(V-I),0}=0.024$,  $\sigma_{A_V,0}=0.048$
and $\sigma_{A_I,0}=0.024$.
To check our maps we present in Fig. \ref{fig:AVSC30-31} a comparison of $A_V$
in the relatively large overlap region of BUL\_SC30 and BUL\_SC31.
One can see a good correlation between them in this figure, and $\chi^2/d.o.f.
= 1.15$ implies that our error estimate is reasonable.

\begin{table*}
 \caption{Basic parameters of maps,
   the bin-size, the number of groups, the total number of all stars, the mean number
   of all stars in each bin, the total number of RCGs, mean of RCGs in each group,
   the mean of values and errors in $E(V-I)$, $A_V$ and $A_I$.
    \label{tbl:AVmaps}}
    \begin{tabular}{rrrrrrcrcrcccccc}\\
    BUL & binsize &  $N_{\rm grp}$
  & \multicolumn{2}{c}{$N_{\rm all}$}
  & \multicolumn{2}{c}{$N_{\rm RC}$}
  & \multicolumn{2}{c}{$E(V-I)$}
  & \multicolumn{2}{c}{$A_V$}
  & \multicolumn{2}{c}{$A_I$}
    \\
    \_SC & (pixel) &
  & total & mean
  & total & mean
  & mean & $\langle \sigma \rangle$ 
  & mean & $\langle \sigma \rangle$ 
  & mean & $\langle \sigma \rangle$ 
  \\
  \hline
 1 &  73.1 &  189 & 647014  & 203.1 & 18571 & 98.3 & 0.854 & 0.012 & 1.677 & 0.027 & 0.823 & 0.017 \\
 2 &  68.3 &  207 & 753366  & 198.4 & 20416 & 98.6 & 0.787 & 0.012 & 1.545 & 0.024 & 0.759 & 0.013 \\
 3 &  73.1 &  427 & 653389  & 205.3 & 42183 & 98.8 & 1.472 & 0.014 & 2.891 & 0.065 & 1.419 & 0.061 \\
 4 &  70.6 &  417 & 693066  & 197.7 & 41059 & 98.5 & 1.319 & 0.014 & 2.591 & 0.054 & 1.272 & 0.048 \\
 5 & 157.5 &    - & 145562  & 209.2 & 21310 & 35.6 & 2.918 & 0.030 & 5.733 & 0.192 & 2.814 & 0.184 \\
 6 & 107.8 &   68 & 319093  & 221.0 & 6783 & 99.8 & 0.700 & 0.011 & 1.374 & 0.024 & 0.675 & 0.015 \\
 7 &  97.5 &   62 & 376688  & 210.3 & 6176 & 99.6 & 0.676 & 0.012 & 1.327 & 0.025 & 0.652 & 0.015 \\
 8 & 113.8 &   52 & 287653  & 214.7 & 4476 & 86.1 & 1.087 & 0.014 & 2.135 & 0.042 & 1.048 & 0.034 \\
 9 & 113.8 &   50 & 269318  & 197.2 & 4516 & 90.3 & 1.060 & 0.013 & 2.083 & 0.039 & 1.023 & 0.032 \\
10 & 107.8 &   61 & 316050  & 219.4 & 5382 & 88.2 & 1.136 & 0.013 & 2.231 & 0.044 & 1.095 & 0.037 \\
11 & 113.8 &   58 & 287074  & 220.4 & 4965 & 85.6 & 1.153 & 0.014 & 2.266 & 0.046 & 1.112 & 0.039 \\
12 &  93.1 &   84 & 400853  & 215.7 & 7154 & 85.2 & 1.165 & 0.015 & 2.289 & 0.048 & 1.123 & 0.040 \\
13 &  89.0 &   83 & 455162  & 212.6 & 7088 & 85.4 & 1.047 & 0.015 & 2.056 & 0.042 & 1.009 & 0.034 \\
14 &  93.1 &  158 & 409086  & 201.4 & 15727 & 99.5 & 1.269 & 0.012 & 2.494 & 0.048 & 1.224 & 0.044 \\
15 & 102.4 &  134 & 322254  & 221.3 & 13232 & 98.8 & 1.408 & 0.012 & 2.766 & 0.059 & 1.358 & 0.055 \\
16 &  89.0 &  127 & 444114  & 232.1 & 11926 & 93.9 & 1.094 & 0.013 & 2.150 & 0.038 & 1.055 & 0.031 \\
17 &  85.3 &  131 & 494911  & 213.3 & 12606 & 96.2 & 0.988 & 0.012 & 1.940 & 0.031 & 0.952 & 0.022 \\
18 &  73.1 &  189 & 649856  & 196.7 & 18267 & 96.7 & 0.904 & 0.012 & 1.776 & 0.028 & 0.872 & 0.019 \\
19 &  75.9 &  173 & 625566  & 203.6 & 16562 & 95.7 & 1.022 & 0.012 & 2.008 & 0.034 & 0.986 & 0.026 \\
20 &  68.3 &  313 & 732180  & 193.8 & 30732 & 98.2 & 0.986 & 0.013 & 1.936 & 0.032 & 0.951 & 0.023 \\
21 &  66.1 &  287 & 793369  & 197.6 & 27964 & 97.4 & 0.930 & 0.013 & 1.827 & 0.030 & 0.897 & 0.020 \\
22 &  75.9 &  273 & 596019  & 194.3 & 26357 & 96.5 & 1.392 & 0.013 & 2.734 & 0.060 & 1.342 & 0.055 \\
23 &  78.8 &  230 & 583168  & 204.6 & 22400 & 97.4 & 1.377 & 0.013 & 2.704 & 0.058 & 1.327 & 0.053 \\
24 &  81.9 &  228 & 509439  & 196.1 & 22737 & 99.7 & 1.282 & 0.011 & 2.519 & 0.049 & 1.237 & 0.045 \\
25 &  81.9 &  210 & 519559  & 198.7 & 20908 & 99.6 & 1.191 & 0.011 & 2.339 & 0.042 & 1.148 & 0.037 \\
26 &  75.9 &  175 & 617829  & 201.2 & 16864 & 96.4 & 0.945 & 0.011 & 1.857 & 0.029 & 0.912 & 0.021 \\
27 &  75.9 &  156 & 594953  & 193.9 & 15333 & 98.3 & 0.860 & 0.011 & 1.690 & 0.025 & 0.830 & 0.016 \\
28 & 102.4 &   80 & 339107  & 201.0 & 7871 & 98.4 & 0.834 & 0.012 & 1.638 & 0.025 & 0.804 & 0.015 \\
29 &  93.1 &   76 & 398252  & 194.7 & 7430 & 97.8 & 0.779 & 0.011 & 1.530 & 0.024 & 0.751 & 0.015 \\
30 &  73.1 &  258 & 671739  & 211.3 & 25476 & 98.7 & 0.971 & 0.012 & 1.908 & 0.030 & 0.937 & 0.022 \\
31 &  70.6 &  245 & 716500  & 204.2 & 24183 & 98.7 & 0.921 & 0.012 & 1.810 & 0.028 & 0.888 & 0.018 \\
32 &  70.6 &  217 & 684274  & 200.0 & 21387 & 98.6 & 0.822 & 0.012 & 1.614 & 0.025 & 0.792 & 0.014 \\
33 &  73.1 &  184 & 669836  & 202.6 & 17962 & 97.6 & 0.865 & 0.012 & 1.699 & 0.026 & 0.834 & 0.016 \\
34 &  64.0 &  331 & 849632  & 198.4 & 31842 & 96.2 & 1.146 & 0.013 & 2.252 & 0.042 & 1.106 & 0.035 \\
35 &  70.6 &  221 & 696335  & 196.7 & 21495 & 97.3 & 0.936 & 0.012 & 1.838 & 0.029 & 0.902 & 0.019 \\
36 &  68.3 &  205 & 764524  & 201.5 & 19973 & 97.4 & 0.825 & 0.012 & 1.621 & 0.026 & 0.796 & 0.015 \\
37 & 102.4 &    - & 463048  & 274.3 & 37114 & 23.4 & 1.921 & 0.029 & 3.773 & 0.114 & 1.852 & 0.102 \\
38 &  73.1 &  213 & 634009  & 199.2 & 21036 & 98.8 & 0.927 & 0.012 & 1.821 & 0.029 & 0.894 & 0.019 \\
39 &  70.6 &  383 & 684042  & 193.9 & 37753 & 98.6 & 1.337 & 0.013 & 2.625 & 0.055 & 1.289 & 0.050 \\
40 &  93.1 &  230 & 419439  & 206.9 & 22680 & 98.6 & 1.496 & 0.012 & 2.940 & 0.066 & 1.443 & 0.063 \\
41 &  85.3 &  229 & 495001  & 205.2 & 22661 & 99.0 & 1.351 & 0.012 & 2.653 & 0.055 & 1.303 & 0.051 \\
42 &  81.9 &  160 & 521529  & 198.2 & 14804 & 92.5 & 1.164 & 0.013 & 2.286 & 0.043 & 1.122 & 0.036 \\
43 & 120.5 &  247 & 257326  & 216.0 & 25921 & 104.9 & 1.870 & 0.012 & 3.674 & 0.097 & 1.803 & 0.094 \\
44 &     - &   -  &  53346  &  -    & 1257  & -    & -     & -     & -     & -     & -     & -     \\
45 &  78.8 &  175 & 562718  & 205.4 & 17247 & 98.5 & 0.836 & 0.012 & 1.642 & 0.028 & 0.806 & 0.018 \\
46 &  85.3 &  159 & 495299  & 213.1 & 15643 & 98.4 & 0.871 & 0.012 & 1.711 & 0.029 & 0.840 & 0.019 \\
47 & 128.0 &   58 & 209811  & 194.4 & 4986 & 86.0 & 1.321 & 0.015 & 2.595 & 0.056 & 1.274 & 0.049 \\
48 & 128.0 &   57 & 217083  & 201.0 & 5229 & 91.7 & 1.194 & 0.014 & 2.346 & 0.046 & 1.152 & 0.039 \\
49 & 136.5 &   43 & 193318  & 203.5 & 4086 & 95.0 & 1.063 & 0.014 & 2.089 & 0.038 & 1.025 & 0.030 \\
\end{tabular}\\
Note:  $\langle \sigma \rangle$ do not
include zero-point errors $\sigma_{E(V-I),0}=0.024$,  $\sigma_{A_V,0}=0.048$
and $\sigma_{A_I,0}=0.024$.
\end{table*}

\newpage

\begin{figure}
\includegraphics[angle=-90,scale=0.37,keepaspectratio]{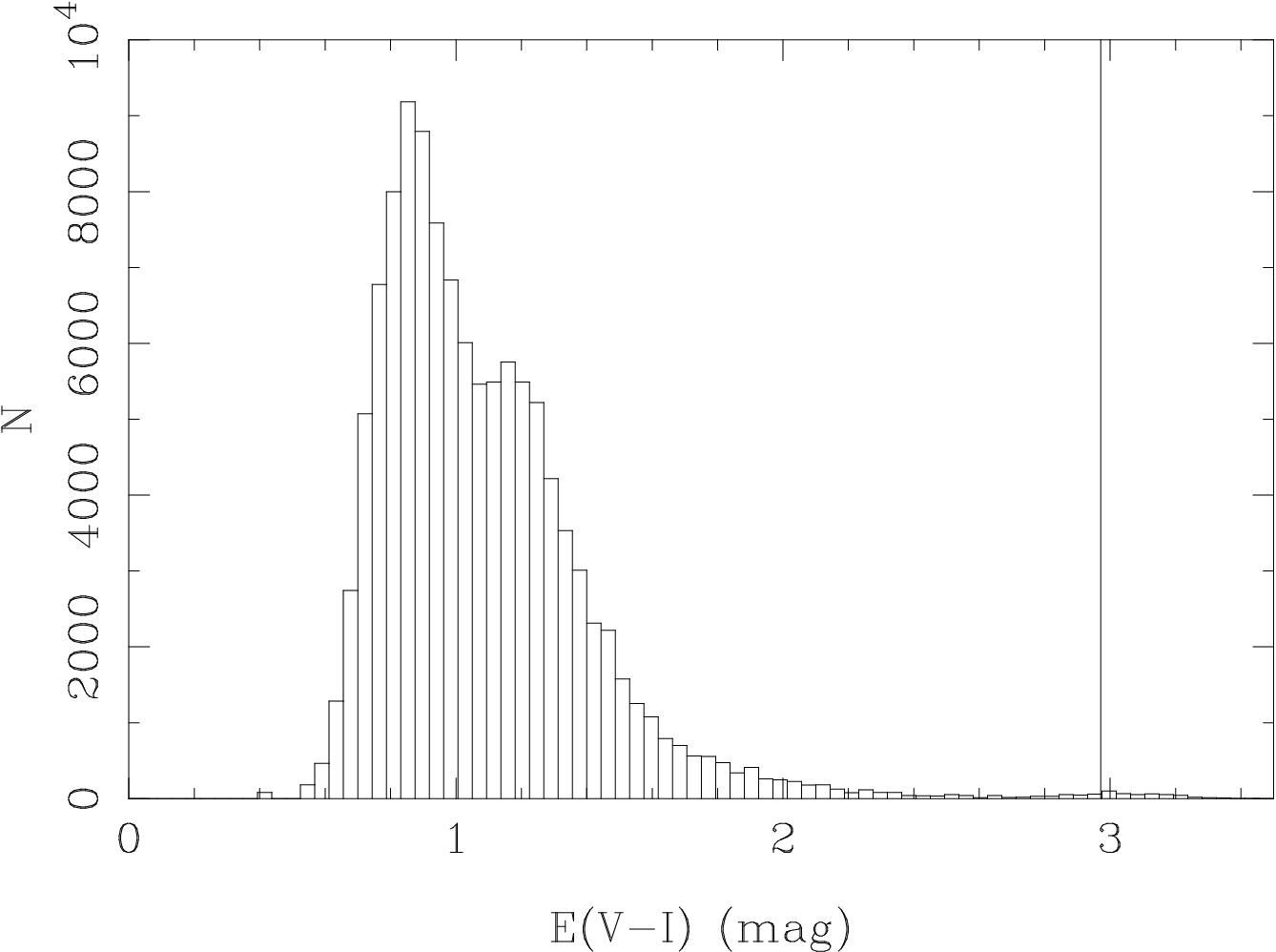}
\begin{center}
\caption{
A histogram of $E(V-I)$ for all our maps. Vertical line indicates the threshold
value $E(V-I)_{\rm th}=3.0$ mag.
\label{fig:hist}
}
\end{center}
\end{figure}

We show the histogram of $E(V-I)$ for all our maps in Fig. \ref{fig:hist}.
The vertical line indicates the threshold value $E(V-I)_{\rm th} = 3.0$, which
corresponds to $\langle V-I \rangle _{\rm RC} = 4$ mag.
We have confidence in our maps below this threshold.  However, small, very
heavily obscured regions are above this reddening threshold, pushing some RCGs
below the OGLE-II detection limit, and distorting the RCGs.
This implies that for $\langle V-I \rangle _{\rm RC} > 4$ mag
our maps give a lower limit to the extinction.
Histograms of $A_V$ and $A_I$ can be obtained by multiplying $E(V-I)$
by $R_{VI}$ and $R_I$ respectively.
The thresholds are: $A_{V, \rm th} = 5.8$ and $A_{I,\rm th} = 2.9$.

The reddening corrected value of $(V-I)_{\rm 0,RC}$ may vary from one
field to another due to weak dependence on metalicity.
The RR Lyrae variables are good distance indicators (e.g., \citealt{nem94})
whose period-luminosity relations are well established (\citealt{jon92}).
The RR Lyrae lie in the instability strip, the range of their colour is small
and it weakly depends on period, amplitude and/or metallicity (\citealt{bon94}; \citealt{alc98a}).
To check the relative zero-points of our reddening maps for each field
we make use of the $V-I$ colour of RR Lyrae Type ab (RRab),
assuming that period - $(V-I)_0$ colour
relation of RRab is the same for all OLGE-II fields

\begin{figure}
\includegraphics[angle=-90,scale=0.37,keepaspectratio]{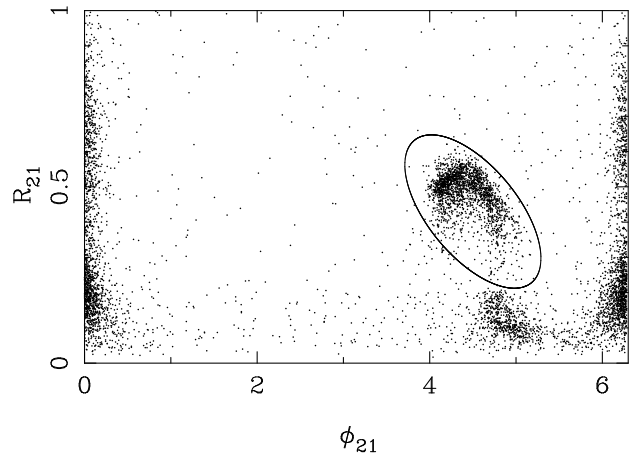}
\begin{center}
\caption{
$R_{21}$ v.s. $\phi_{21}$ for periodic variables in the OGLE catalogue.
RRab variables are selected within the ellipse.  There is a clump
of RRc variables below the ellipse, but we do not use them in this work. 
\label{fig:R21}
}
\end{center}
\end{figure}

\begin{figure}
\includegraphics[angle=0,scale=0.4,keepaspectratio]{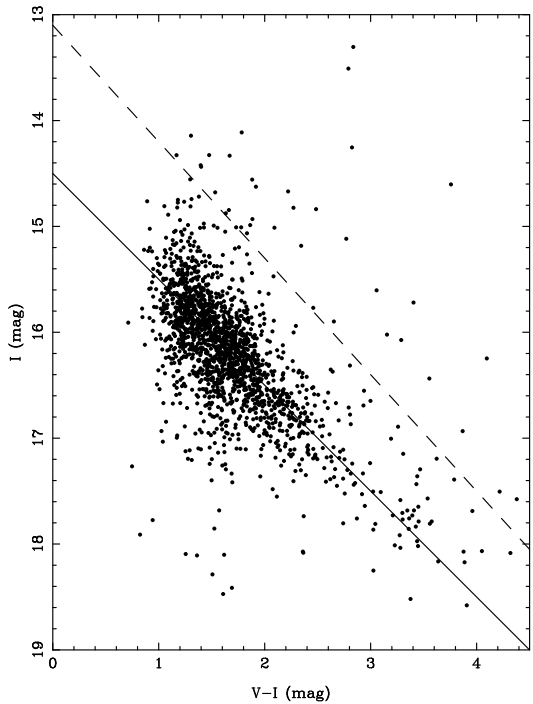}
\begin{center}
\caption{
The CMD for RRab variables. Solid line with a slope of $I/(V-I)=1.0$ is 
shown as a reference. RRab variables above the dashed line: $I=1.1(V-I)+13.1$
are rejected in our analysis because they are foreground disk stars
or variables blended with other bright stars.
\label{fig:CMDRRab}
}
\end{center}
\end{figure}

We selected RRab in OGLE-II Galactic Bulge
variable star catalogue (\citealt{woz01}) by using the method of 
\cite{ala96}.  We measured periods $P$ using PDM algorithm (\citealt{ste78})
and Fourier coefficients by fitting Fourier series with five harmonics.
In Fig \ref{fig:R21} we show the the amplitude ratio $R_{21}\equiv A_2/A_1$
versus phase differences $\phi_{21}\equiv \phi_2-2\phi_1$ for
variables with $P<0.9$ days,
except $0.4985 <P<0.5001$ days, which are affected by aliasing. 
We selected 1,961 RRab stars as a clear clump within the ellipse 
in this figure.  Of these 1,819 stars have $ \langle V \rangle $ and
$ \langle I \rangle $ photometry provided by \cite{uda03}.
We visually inspected all light curves and several stars with non RR Lyrae 
shape light curves were rejected.  In Fig \ref{fig:CMDRRab}
we present colour - magnitude diagram for RRab stars so selected.
The differential reddening is apparent.  A similar reddening
slope is found in \S \ref{sec:AIEVI}.  

In the following analysis we rejected 39 stars with $I<1.1(V-I)+13.1$
because they are either nearby disk stars or they are
blended with other bright stars.  We did not reject several background RRab
below the sequence as they made no difference
to our analysis, for which we used 1,780 RRab stars.

We assume that reddening corrected colour
$(V-I)_{\rm 0,RR}$ is the same in every field, with a possible
weak dependence on period $P$, or amplitude $A$, or both.
We estimate the zero-points $C_i$ for each 
field $i$ by fitting extinction corrected colour of RRab with a linear
dependence on $\log P$, $A$, and on both of them.  We obtained the
following relations:
\begin{eqnarray}
  \label{eq:VIP} 
  (V-I)_{\rm 0,RR} &=& 0.309 (\pm 0.049) {\rm log}P + C_{i}, sdev= 0.158,\\
  \label{eq:VIA} 
  (V-I)_{\rm 0,RR} &=& -0.096 (\pm 0.016) A + C_i, sdev= 0.157,\\
  \label{eq:VIPA} 
  (V-I)_{\rm 0,RR} &=& 0.211 (\pm 0.056) {\rm log}P -0.063 (\pm 0.018) A   \nonumber \\
                 &+& C_i, sdev= 0.156.  
\end{eqnarray}
All three relations provided almost the same values of $C_i$. 
Fitting with both $P$ and $A$ seems to be an over-parameterization.

\begin{figure}
\includegraphics[angle=0,scale=0.5,keepaspectratio]{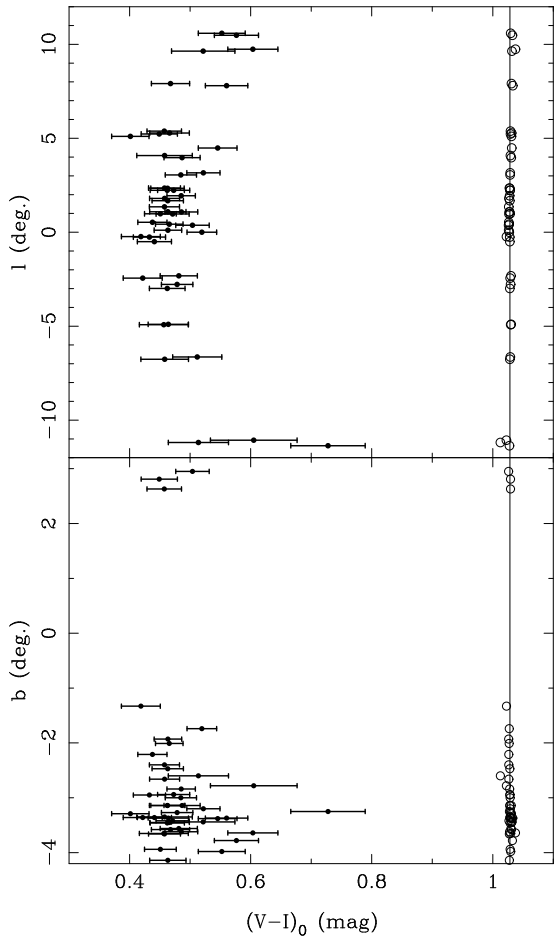}
\begin{center}
\caption{
Extinction corrected colours of RCGs $(V-I)_{\rm 0,RC}$ (open circle) and RRab
$(V-I)_{\rm 0,RR}$  (filled circle and errors) as a function of the galactic 
$l$ and $b$. $(V-I)_{\rm 0,RR}$ are the value  given by Eq. (\ref{eq:VIP})
at log$P=-0.3$.  Solid line represent a mean of 
$(V-I)_{\rm 0,RC}= 1.0283 \pm 0.024, sdev=0.0034$,
where the error in $(V-I)_{\rm 0,RC}$ is dominated by 
$\sigma_{E(V-I),0}=0.024$.
\label{fig:VIRCc}
}
\end{center}
\end{figure}

In Fig. \ref{fig:VIRCc}, we show extinction corrected colours of RCGs 
$(V-I)_{\rm 0,RC}$ (open circle) and RRab $(V-I)_{\rm 0,RR}$ 
(filled circle and errors) as a function of the galactic $l$ and $b$, 
where $(V-I)_{\rm 0,RC}$ are measured by the method [3] in 
\S\,\ref{sec:AIEVI} and
$(V-I)_{\rm 0,RR}$ are the value given by Eq. (\ref{eq:VIP}) at log$P=-0.3$.

$(V-I)_{\rm 0,RC}$ is constant as assumed.
The mean of $(V-I)_{\rm 0,RC}= 1.0283 \pm 0.024$,
where we used $\sigma_{E(V-I),0}=0.024$ as the error in $(V-I)_{\rm 0,RC}$,
is consistent with the colour of nearby RCGs
$(V-I)_{\rm 0,RC,near}= 1.01\pm 0.08$ mag (\citealt{pac98}), contrary to
$(V-I)_{\rm 0,RC,bulge}= 1.11\pm 0.12$ by \cite{pac99} with \cite{sta96}'s
map.

\begin{figure}
\includegraphics[angle=-90,scale=0.37,keepaspectratio]{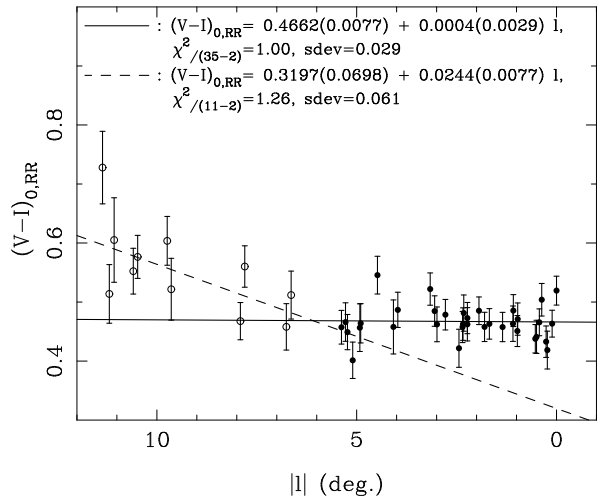}
\begin{center}
\caption{
Extinction corrected colours of RRab $(V-I)_{\rm 0,RR}$ as a function of 
the absolute galactic $|l|$.
$(V-I)_{\rm 0,RR}$ are the value  given by Eq. (\ref{eq:VIP}) at log$P=-0.3$.
Filled and open circles represent plots in $|l|<6$ and in $|l|>6$, 
and solid and dashed lines indicate the best fits for them, respectively.
One can see the RRab colour is constant at $|l|<6$ and significantly red at $|l|>6$.
\label{fig:VIRRc}
}
\end{center}
\end{figure}

$(V-I)_{\rm 0,RR}$ seems to be systematically redder by about 
$\sim 0.1$ mag at large $|l|$.  Such trend can not be seen in $b$.  
We plot $(V-I)_{\rm 0,RR}$ as a function $|l|$ in Fig. \ref{fig:VIRRc}
and we fit it with a straight line for $|l|<6$ (filled circle and solid
line) and for $|l|>6$ (open circle and dashed line).
One can see the RRab colour is constant at $|l|<6$ and have 
a significant ($3\sigma$) dependence on $|l|$ at $|l|>6$.
The mean $(V-I)_{\rm 0,RR}$ is $0.4670 \pm0.0047$ with $sdev=0.0286$
for $|l|<6$ and  $0.5386 \pm0.0126$ with $sdev=0.0773$ for $|l|>6$.
The significance of the redder colour in  $|l|>6$ is estimated to be 
$5.3 \sigma$ level.

This implies that
$(V-I)_{\rm 0,RC}$, $(V-I)_{\rm 0,RR}$ or both of them, may vary with $l$.
If we assume that $(V-I)_{\rm 0,RR}$ is constant, then the intrinsic colour of
RCGs would be bluer, i.e., the zero-points of our reddening maps would be 
smaller by $\sim 0.1$ mag at $|l|=10$, compared to $|l| \le 6$.
This could be explained if the outer RCGs have lower metalicity and
are somewhat bluer than near the Galactic Center.
The colour of RRab $(V-I)_{\rm 0,RR} = 0.4 \sim 0.5 $ for  $|l| \le 6$ is
consistent with 
the colour of local RRab, while the two sets of colours were claimed
to be different by \cite{sta96}'s map by \citealt{stu99}.

The scatter of $(V-I)_{\rm 0,RR}$ values is large
because public domain $V$-band OGLE photometry is the average
of randomly distributed small number of $V$-band measurements,
while the $V$ amplitude is large.
Furthermore the scatter is larger for  $|l|>6$ because the number 
of RRab is small in these fields.
Therefore, we do not correct for the offsets apparent in Fig. \ref{fig:VIRCc}.
It will be possible to improve the accuracy of $(V-I)_{\rm 0,RR}$ 
when individual $V$-band measurements become available.

\section{Discussion and Conclusion}

\label{sec:disc}

We confirmed the anomalous reddening, i.e. a small value of the ratio of 
total to selective extinction $R_{VI}=1.9 \sim 2.1$ depending on the line
of sight, as measured by \cite{uda03}. This implies that the distribution
of dust grains may be tipped to smaller sizes in these regions, compared
to average in the Galaxy.
A detailed analysis of these results is beyond the scope of the present study.

By adopting the mean value of $R_{VI}=1.964$,
we have constructed reddening $E(V-I)$ and extinctions, $A_V$ and $A_I$ maps
in 48 OGLE-II GB fields, covering a range of $-11^\circ <l< 11^\circ$,
with the total area close to 11 square degrees.
The  reddening $E(V-I)$ and extinctions, $A_V$ and $A_I$ are measured in the 
range $0.42<E(V-I)<3.5$ mag, $0.83<A_V<6.9$ mag and $0.41<A_I<3.4$.
Note: above the threshold values $E(V-I)_{\rm th} = 3.0$, $A_{V, \rm th}=5.8$ 
and $A_{I,\rm th} = 2.9$, which correspond to 
$\langle V-I \rangle _{\rm RC} = 4$ mag, our maps give the lower limit to
reddening and extinction.  Spatial resolutions of maps are $26.7''\sim 106.8''$ 
depending on the stellar density of each field.

The absolute zero-point is calibrated using 20 RRab variables in Baade's
Window, following \cite{alc98b}.
Relative zero-points of our maps are verified with 1,780 RRab variables
found in our fields.  We found that these zero-points may be lower
by $\sim 0.1$ mag at larger Galactic longitudes $|l|>6$.
Note that our extinction map is OK in terms of the relative extinction
within each field. The relative zero points are also OK between fields 
for $|l|<6$. We did not make any correction for this effect in the paper.

We used the mean value of $R_{VI}=1.964$, but different fields have a range
of values, with a standard deviation $sdev \sim 0.085$ (see Fig.\ref{fig:RIlb}).
We estimated the errors of our maps by taking this range into account.
The errors approach $0.17$ mag at the highest extinction ($A_V=6$).

As noted by \cite{uda02} and \cite{uda03}, the $I$-band filter used by OGLE-II
has the red wing somewhat wider than standard.  This
may lead to systematic deviations from the 
standard values for very red stars,
giving brighter $I$-band magnitudes (redder $V-I$ colour)
for the OGLE-II filter for very red stars ($V-I>2$), while the error is
negligible in the range
where the OGLE-II data were calibrated by standards, i.e. for ($V-I<2$). 
This effect may reach $\sim 0.2$ mag for the very red RCGs ($V-I>4$ mag).
This effect makes the slope of reddening line in the CMD somewhat shallower,
as shown by \cite{uda03}. This effect is not sufficient to explain the 
anomalous extinction towards the GC.
\cite{uda03} analyzed the differences between OGLE-II and standard 
$I$-band filters using \cite{kur92}'s model
atmosphere of a typical RCGs, reddened with the standard interstellar 
extinction of \cite{car89} and \cite{fit99}. 
He found possible errors to be at the level
$ \pm 0.1$ in $R_{VI}$, depending on the 
model of extinction and the colour range of RCGs in each field. 
This leads to $\sim 0.2$ mag differences in $E(V-I)$, $A_V$ and $A_I$ 
for the very red
RCGs, with $V-I>4$ mag, corresponding to $E(V-I) \sim 3$, $A_V\sim 5.9$ and 
$A_I=2.9$ mag in our extinction maps.
These differences are the upper limits because the atmospheric absorption
in the range 900-990 nm makes the OGLE-II $I$-band filter
closer to the standard one (\citealt{uda02}).
To make a more accurate estimate of the required correction it
is necessary to have very red standard stars, with $(V-I) > 2 $.

\begin{figure*}
\includegraphics[angle=-90,scale=0.65,keepaspectratio]{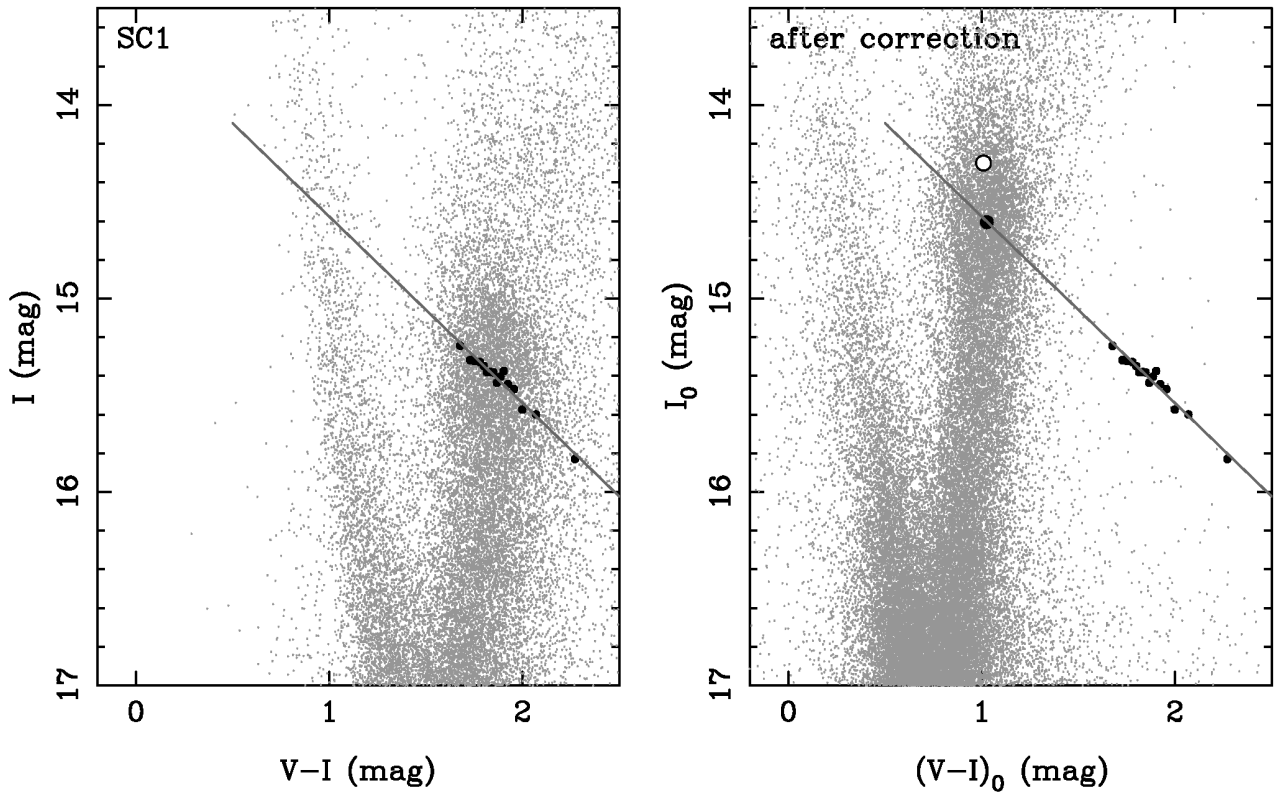}
\begin{center}
\caption{
Colour Magnitude Diagram of BUL\_SC1 before (left panel) and after
(right panel) the extinction and reddening correction.
Small and large filled circles indicate the same plots as
Fig. \ref{fig:IVI} and RCGs centroid after the extinction and
reddening correction.
Open circle represents the RCGs centroid expected from the assumption
that the population of RCGs  in GB is the same as local.
The solid line indicate the reddening line given by Eq.
\ref{eq:reddningcurve} with the mean slope $ R_I = 0.964 $ and a
constant $I_{0,RC}=14.6$ for this field.
\label{fig:CMDRCc}
}
\end{center}
\end{figure*}

The extinction corrected $I$-band magnitude of RCGs, $I_{0,RC}^{*}=14.6$
in Baade's window is fainter than expected. Adopting
distance modulus to the GC of $14.52\pm0.1$ mag (\citealt{eis03}), and
assuming that the population of RCG stars in the Galactic bulge is
the same as local, i.e. that the absolute magnitude is
$M_{I_{0,RC}}=-0.26\pm 0.03$ (\citealt{alv02}) and the average colour is
$(V-I)_0 = 1.01\pm 0.08$ (\citealt{pac98}), then the expected
magnitude of RCG in Baade's Window should be $I_{0,RC} = 14.3$.
In Fig. \ref{fig:CMDRCc} we show CMD and RCG centroids of BUL\_SC1 
(at Baade's window) before (left panel) and after (right panel) the 
extinction and reddening correction, with the expected RCG centroid.
Note, that the difference between the average distance modulus in
Baade's Window and the Galactic center is only 0.02 mag, i.e. it is
of no consequence (\citealt{pac98}).

We do not know what is the solution of this problem.  It may be that the 
population effects are large $M_{I_{0,RC}}=-0.4 \sim -0.03$, as claimed by
\cite{sal03} and \cite{per03}, it
may be that the distance to the GC is larger, or it is possible that
the reddening is more complicated.  We assumed that the extinction
to reddening ratio is constant all the way to zero extinction.  OGLE 
photometry is well calibrated with standards for $(V-I) < 2$.  The slope 
of the reddening line in Baade's Window (Region D) is well measured
for RCG in the colour range $ 1.6 < (V-I) < 2.2 $, but we have no direct
information about the reddening line for $ 1.0 < (V-I) < 1.6 $, i.e.
for the reddening range $ 0 < E_{V-I} < 0.6 $.  If we make an ad hoc
assumption that the RCG population in Baade's Window is the same as
local, and the distance is 8 kpc, then we may obtain
$I_{0,RC}^{*}=14.3$ and $(V-I)_{0,RC}=1.0$ adopting
$R_{VI} \sim 2.8$ (i.e. $R_{I} \sim 1.8$)for the unobserved range
$ 0 < E_{V-I} < 0.6 $. 
We do not know if this is plausible or not.

The situation will improve somewhat once detailed $V$-band OGLE photometry
becomes available for RR Lyrae stars, and this will make it possible to 
obtain an independent estimate of the reddening in Baade's Window.
Preliminary analysis of the average photometry of RR Lyrae stars seems
to indicate that the puzzling brightness of red clump giants may be due
to population effects, i.e. their average absolute magnitude in the Bulge
is somewhat different than it is near the Sun.  A much improved analysis
will be done when OGLE-II $V$-band measurements
become available in the next several months.
At this time it should be OK to use our maps of differential reddening,
but we consider our calibration of zero point to be uncertain.  This
implies that at this time our reddening maps are not adequate for
a quantitative study of Galactic bar structure.

In this work we leave our extinction maps in the OGLE-II $I$-band because 
the reddening map is urgently needed for various applications, 
and is already used in some works (\citealt{sum03b}; \citealt{war03}).
A reader must take care of possible small error mentioned above while
using our maps for standard $I$-band photometry.

The extinction map is available in electronic format via
anonymous ftp from the server 
{\it ftp://ftp.astrouw.edu.pl/ogle/ogle2/extinction/} and
{\it ftp://bulge.princeton.edu/ogle/ogle2/extinction/}.

These extinction maps of OGLE-II GB fields facilitate the study of the 
Galactic structure with OGLE proper motion catalogue (\citealt{sum03b})
and microlensing optical depth, and a study of variable stars,
but the reader should be aware of the zero point of the extinction
may not be accurate.  We intend to improve the quality of the zero
points in all fields as soon as individual OGLE $V$-band measurements of
the RR Lyrae stars become available.

\section*{Acknowledgments}

We are grateful to B. Paczy\'{n}ski for helpful comments and discussions. 
We would like to thank  A.~Udalski for important suggestions.
We acknowledge B.~Drain, D.~Schlegel and D.~Finkbeiner for 
carefully reading the manuscript and comments.
T.S. acknowledge the financial support from the JSPS.
This work was partly supported with the following grants to B. Paczy\'nski:
NSF grant AST-0204908, and NASA grant NAG5-12212.

\label{lastpage}
\clearpage

\end{document}